\documentclass[twocolumn,prb,showpacs]{revtex4}
\usepackage{graphicx}
\usepackage{psfrag}
\usepackage{color}
\usepackage{subfigure}
\usepackage{amsmath}

\begin{document}
\title{Numerical study of magnetic and pairing correlation in bilayer triangular lattice}
\author{Shuang Wu$^{1}$, Jinling Li$^{1}$, Pan Gao$^{1}$,
Ying Liang$^{1}$, Tianxing Ma$^{1,2,}$\footnote{txma@bnu.edu.cn}}
\affiliation{$^{1}$Department of Physics, Beijing Normal University, Beijing 100875, China%
\\
$^{2}$Beijing Computational Science Research Center, Beijing 100084, China}
\date{\today}
%\maketitle
\begin{abstract}
%With
By using the determinant Quantum Monte Carlo method, the  magnetic and pairing correlation of the
Na$_{x}$CoO$_{2}\cdot$yH$_{2}$O system are studied within the Hubbard model on a bilayer triangular lattice.
%the magnetic correlation and pairing susceptibility in the Hubbard model on a bilayer triangular lattice is studied.
The temperature dependence of spin correlation function and pairing susceptibility with several kinds of symmetries at different
electron fillings and inter layer coupling terms are investigated.
It is found that the system shows an antiferromagnetic correlation around the half filling,
and the $fn$-wave pairing correlation dominates over other kinds of pairing symmetry in the low doping region.
As the electron filling decreases away from the half filling, both the ferromagnetic correlation and the $f$-wave paring susceptibility
are enhanced and tend to dominate. It is also shown that both the magnetic susceptibility and paring susceptibility decrease as the inter layer coupling increases. %It is
%found that the $f$-wave pairing maybe dominate over pairings with
%other symmetries.
\end{abstract}


 \maketitle

\section{Introduction}
Understanding the competition between various magnetic orders and pairing symmetries is a major challenge in superconductivity now. The discovery of superconductivity in the Na$_{x}$CoO$_{2}\cdot$
yH$_{2}$O materials
%and related fascinating normal-state properties has
%attracted much attention\cite{Takada2003}, which
%has set a new milestone to search for new layered metal-oxide superconductors \cite{Takada2003}, and also
offer an appealing platform to investigate the interplay among pairing interactions, magnetic fluctuations, and electronic correlation\cite{Takada2003}.
 %enormous properties in $\text{Na}_{x}\text{Co}\text{O}_{2}\cdot\text{yH}_{2}\text{O}$ have been studied and reported.
Besides doped cuprates, cobaltates
are another class of a layered 3$d$ transition-metal oxide in which the superconductivity
has been observed.
%In the high-Tc copper oxides, superconductivity
%occurs in the CuO_{2} square lattice. The Cu$^{2+}$ moments are antiferromagnetically ordered in the CuO_{2} plane.
%With a low level of carrier (hole or electron) doping, the antiferromagnetism  is suppressed drastically, and the system becomes
%metallic, followed by the appearance of superconductivity.
The main difference between the two systems is that Co ions form a triangular
lattice with magnetically frustrated geometry in contrast to the
square lattice of the CuO$_{2}$ plane.
In doped cuprates, it has been well established that
the Cu$^{2+}$ moments are antiferromagnetically ordered in the CuO$_{2}$ plane,
and with a low level of carrier (hole or electron) doping, the antiferromagnetism
is suppressed drastically, and the system becomes metallic, followed by
the appearance of superconductivity where the d$_{x^{2}-y^{2}}$ pairing symmetry dominates in
the optimally doped region\cite{Kastner1998,Lee2006,Tsuei2000,Yeh2001,Feng2003}. In
general, the doping dependence of the pairing symmetry
and the issue of quantum criticality must be considered
under the premise of spatial homogeneity in the pairing
potential. Results of some experiments suggest triplet pairing in
cobaltates\cite{Higemoto2004,Maska2004}, while some other measurements have resulted in contradicting
conclusions which indicate singlet pairing\cite{Zheng2006,Fujimoto2004}.

% The
%%mechanism of superconductivity in those systems has been discussed in
%several theoretical papers\cite{Baskaran2003,Tanaka2003,Kumar2003,Honerkamp2003,Wang2004,Kuroki2004,Johannes2004,Kuroki2005} but no final conclusions regarding the symmetry and the total spin of the paired state
%have been drawn. On the other hand, the interplay of geometrical frustration and strong electron
%correlation is one of the hot topics in the field of
%electron correlated systems.

To investigate the superconducting mechanism of Na$_{x}$CoO$_{2}\cdot$
yH$_{2}$O system, the triangular lattice has been extensively studied theoritically\cite{Baskaran2003,Tanaka2003,Kumar2003,Honerkamp2003,Wang2004,Kuroki2004,Johannes2004,Kuroki2005,Koretsune2005,Su2008}.
%but no final conclusions regarding the symmetry and the total spin of the paired state
%have been drawn for those system.
In the Hubbard type model for this frustrated system, perturbation theory shows that $d$-wave and $p$-wave superconducting states are stable in hole doped region\cite{Nisikawa2002},
while the renormalization group approach suggests the $d+id$-wave pairing symmetry in the case with the
antiferromagnetic exchange interactions\cite{Honerkamp2003}.
In the strong-coupling Hubbard model or in its strong coupling limits, the $t$-$J$ model,
mean field results again support the $d+id$-wave superconductivity near the half-filling\cite{Baskaran2003,Kumar2003,Wang2004}, which has been confirmed by the variational Monte Carlo study\cite{Watanabe2004},
while in the low density region where the Fermi surface is detached, $f$-wave pairing is proposed to be realized\cite{Kumar2003}.
% The
%%mechanism of superconductivity in those systems has been discussed in
%several theoretical papers\cite{Baskaran2003,Tanaka2003,Kumar2003,Honerkamp2003,Wang2004,Kuroki2004,Johannes2004,Kuroki2005} but
Then, %there are no final conclusions regarding the symmetry and the total spin of the paired state
%have been drawn for those system.
the results obtained above are still actively debated
because they are very sensitive to the approximation used, exact
numerical results are highly desirable for they provide unbiased information and
would serve as useful bench marks for analytical approach.
Moreover, understanding of the magnetic order and
pair symmetries of frustrated system are still missing.
For example, the situation of the
pairing symmetry in $\kappa$-(ET)$_{2}$X is complicated owing to the
existence of frustration\cite{Clay2012}. In a frustrated quantum antiferromagnet, the
introduction of doping with mobile charge carriers may result in the appearance
of unconventional superconductivity\cite{Gan2006}. Bilayer triangular lattice is an idea platform to study the
interplay between magnetic fluctuation and pairing correlation in frustrated system.

%The complexity of these family members calls for further studies on
%the anisotropic triangular lattice so to clarify some important
%issues.

%About the superconducting properties on the triangular lattice

Again similar to the doped cuprates, Na$_{x}$CoO$_{2}\cdot$yH$_{2}$O are layered
materials, where the distance and couplings between the two
$\text{Co}\text{O}_{2}$ layers depend on the $\text{ H}_{2}\text{O}$ molecules
inserted \cite{Lynn2003,Jorgensen2003,Johannes2004B,Xiao2006},
%, which is of great interest from both chemical and physical point of
%view.
and the inter layer coupling term is also regarded as a key to understand the superconducting mechanism.
%To study the influences of $\text{
%H}_{2}\text{O}$ molecules between the two $\text{Co}\text{O}_{2}$ layers, it is
%of great interest to study the single-band Hubbard model on the bilayer
%triangular lattices\cite{Hu2009}.
Thus, in this paper, we study the magnetic and pairing correlation within the Hubbard model on a bilayer triangular lattice by using
the determinant Quantum Monte Carlo simulations, which is a method that do not rely on uncontrolled approximations\cite{Hu2009,Blankenbecler1981,Ma2010,MaPRL2013,MaReview2011}.
Numerical calculation reported here include results for a variety of band fillings, temperatures, pairing symmetries, and inter layer coupling terms.
It is found that the system shows an antiferromagnetic correlation around the half filling, and the $fn$-wave pairing correlation dominates over other kinds of pairing symmetry in low doping region.
As the electron filling decreases, both the ferromagnetic fluctuations and the $f$-wave paring susceptibility are enhanced and tend to dominate.
It is also shown that the magnetic correlation and paring susceptibility decreases as the inter layer coupling increases.
These results indicate that the competition of ferromagnetic and antiferromagnetic fluctuations in different filling region is crucial on the pairing behavior, which could be understood
from the shape of the density of state (DOS) distribution in bilayer triangular lattice.

\section{Model}

\begin{figure}[tbp]
\includegraphics[scale=0.45]{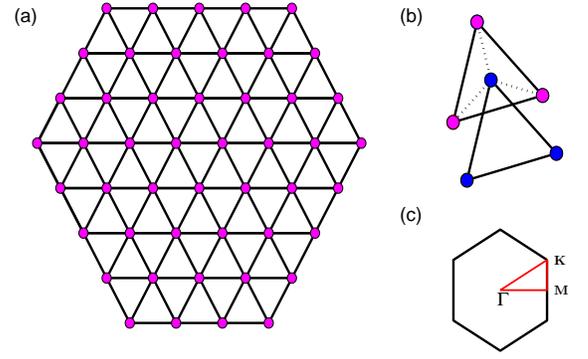}
\caption{(Color online) (a) Sketch of the triangular lattice, (b) bilayer triangular lattice structure and (c) the first Brillouin zone. The red line represent the high symmetry points including $\Gamma$, $M$ and $K$ points.} \label{Fig:Sketch}
\end{figure}

%{\color{blue}
The sketch for the bilayer triangular lattice has been shown in Figs. \ref{Fig:Sketch} (a) and (b).
As shown in Fig. 1(a), the model for each layer is set on a triangular lattice
with hexagonal shape. There are 2N sites on the diagonal,
and the site number of this series of lattice is 3N$^{2}$. This
lattice setting reserves most geometric symmetries of the triangular
lattice. Fig. 1(b) indicates the sketch for the interlayer hoping, and hence the total sites for such bilayer triangular lattice is 2$\times$3N$^{2}$. The case of $N=4$ is shown here. %}
The data points in the first Brillouin zone
(BZ) include all the high symmetry points such as $\Gamma$, M, and
K, are shown in Fig. 1(c). For any atom, it has six nearest neighbor atoms in the same layer and three in the other layer, which could be described as
\begin{eqnarray}
H &=&t\sum_{\mathbf{\langle i,j\rangle }d\sigma
}(c_{\mathbf{i}d\sigma }^{\dag }c_{\mathbf{j}d\sigma }+h.c.)
+t^{\prime }\sum_{\mathbf{\langle i,j\rangle }\sigma }(c_{\mathbf{i}%
1\sigma }^{\dag }c_{\mathbf{j}2\sigma }+h.c.) \notag \\
&&+U\sum_{\mathbf{i}d}n_{\mathbf{i}d\uparrow}n_{\mathbf{i}d\downarrow}
 -\mu
\sum_{\mathbf{i}d\sigma }n_{\mathbf{i}d\sigma }
\end{eqnarray}
where $c_{\mathbf{i}d\sigma}$ ($c_{\mathbf{i}d\sigma}^{\dagger}$) annihilates
(creates) electrons at the site $R_{\mathbf{i}}$ in the $d$-th layer ($d=1,2$)
with spin $\sigma$ ($\sigma=\uparrow,\downarrow$) and
$n_{\mathbf{i}d\sigma}=c^{\dagger}_{\mathbf{i}d\sigma}c_{\mathbf{i}d\sigma}$.
This system has intra layer nearest neighbor hopping $t$ and inter layer
hopping term $t^{\prime}$, and these two layers have the same
chemical potential $\mu$, as well as the electron-electron Coulomb
interaction $U$.
%{\color{blue}
The system was simulated using determinant quantum Monte Carlo at finite temperature, and our numerical calculations were mainly performed on a $2\times 48$ (N=4),  $2\times 75$(N=5) and  $2\times 108$ (N=6) lattices with periodic boundary conditions.
%}
%In such a bilayer triangu.
%The basic strategy of DQMC is to express the partition function as a high-dimensional integral over a set of
%random auxiliary fields. The integral is then accomplished by
%Monte Carlo techniques. In our simulations, 8000 sweeps were used to equilibrate the system. An
%additional 30000 sweeps were then made, each of which generated a
%measurement. These measurements were split into ten bins which provide the
%basis of coarse-grain averages and errors were estimated based on standard
%deviations from the average. For more technique details we refer to
%Refs.~\cite{Blankenbecler1981,MaReview2011}.

\begin{figure}[tbp]
\includegraphics[scale=0.45]{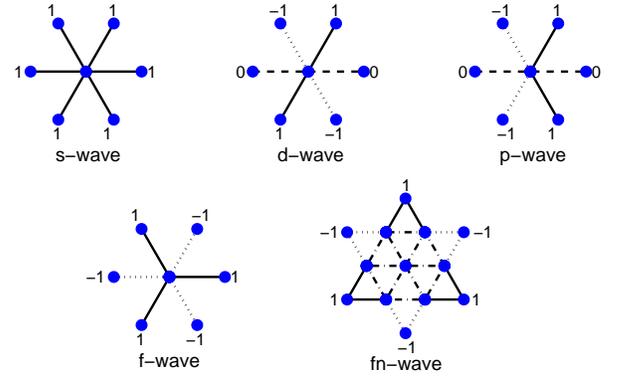}
\caption{(Color online) Site-dependent form factors for $s$ wave, $d$ wave, $p$ wave, $f$ wave, and $fn$ wave pairing correlation functions in the triangular lattice.}
\label{Fig:Pairing}
\end{figure}
It is generally believed that magnetic excitation might play a fundamental role in the superconducting mechanism of electronic correlated systems.
%, especially the antiferromagnetic correlation.
To study the magnetic properties,
%we define the
%spin correlation in the $z$ direction at zero frequency,
%\begin{eqnarray}
%\chi(q) = \int_{0}^{\beta}d\tau \sum_{d,d'=a,b} \sum_{i,j}
%e^{iq\cdot(i_{d}-j_{d'})} \langle\textrm{m}_{i_{d}}(\tau) \cdot
%\textrm{m}_{j_{d'}}(0)\rangle
%\end{eqnarray}
%and
we define the spin susceptibility in the $z$ direction at zero frequency,
\begin{eqnarray}
\chi(q) = \int_{0}^{\beta}d\tau \sum_{d,d'=1,2} \sum_{\mathbf{i},\mathbf{j}} e^{\mathbf{i}q\cdot({\mathbf{i}_{d}}-{\mathbf{j}_{d^{\prime}}})}\langle m_{\mathbf{i}d}\cdot m_{\mathbf{j}d^{\prime}}\rangle,
\end{eqnarray}
where $m_{\mathbf{i}d}(\tau)=e^{H\tau}m_{\mathbf{i}d}(0)e^{-H\tau}$,
$m_{\mathbf{i}d}=c_{\mathbf{i}d\uparrow}^{\dagger}c_{\mathbf{i}d\uparrow}-c_{\mathbf{i}d\downarrow}^{\dagger}c_{\mathbf{i}d\downarrow}$,
%$m_{\mathbf{i}2}=c_{\mathbf{i}2\uparrow}^{\dagger}c_{\mathbf{i}2\uparrow}-c_{\mathbf{i}2\downarrow}^{\dagger}c_{\mathbf{i}2\downarrow}$,
and $N_{s}$ represents the unit number of the lattice.
%where $m_{i_{a}}(\tau)$=$e^{H\tau} m_{i_{a}}(0) e^{-H\tau}$ with
%$m_{i_{a}}$=$a^{\dag}_{i\uparrow}a_{i\uparrow}-a^{\dag}_{i\downarrow}a_{i\downarrow}$
%and
%$m_{i_{b}}$=$b^{\dag}_{i\uparrow}b_{i\uparrow}-b^{\dag}_{i\downarrow}b_{i\downarrow}$.
%We measure $\chi$ in unit of $\mid$t$\mid$$^{-1}$.
%Hence we study the behavior of magnetic correlation firstly, and the
%results of spin
%To study the pairing properties of the bilayer triangular lattice, % of t sodium cobalt oxide,
%The property of superconducting paring in the bilayer triangular lattice is governed by the behavior of pairing susceptibility, which is defined as
To understand the superconductivity in Na$_{x}$CoO$_{2}\cdot$
yH$_{2}$O materials, the behavior of pairing  % a play very important role.
is one of key issues. The property of pairing could be governed by the pairing susceptibility at zero frequency, which is defined as
\begin{eqnarray}
P_{\alpha}&=&\frac{1}{N_{s}}\sum_{\mathbf{i},\mathbf{j}}\int_{0}^{\beta }d\tau\langle\Delta_{\alpha}^{\dagger}({\mathbf{i},\tau})\Delta_{\alpha}({\mathbf{j},0})\rangle\text{,}
\end{eqnarray}
and
\begin{eqnarray}
\Delta_{\alpha}(\tau)&=&\frac{1}{\sqrt{N_{s}}}\sum_{\mathbf{i}}\Delta_{\alpha}({\mathbf{i}},\tau) \\
\notag
&=&\frac{1}{\sqrt{N_{s}}}\sum_{\mathbf{i},l}f_{\alpha}(l)\langle c_{\mathbf{i}\uparrow}(\tau)c_{\mathbf{i}+l\downarrow}(\tau)\pm c_{\mathbf{i}+l\uparrow}(\tau)c_{\mathbf{i}\downarrow}(\tau)\rangle \notag ,
\end{eqnarray}
where $\alpha$ denotes the symmetry of the pairing function, $i$ is the lattice site, $l$ indicates the neighboring sites, and $f_{\alpha}(l)$ is the site-dependent form factor of electron pairs. Considering the symmetry of the triangular lattice, possible form factors include the six types: $f_{s}(l)$, $f_{d_{xy}}(l)$, $f_{d_{x^{2}-y^{2}}}(l)$, $f_{f}(l)$, $f_{p_{x}}(l)$, and $f_{p_{y}}(l)$.
%You can see their detailed definition in Ref.\cite{Koretsune2005}.
The detail forms of these pairing symmetries have been discussed by T. Koretsune and M. Ogata in Ref. \cite{Koretsune2005}. Following them, the possible form factors of the pairing
correlation functions in the triangular lattice have been shown in Fig. \ref{Fig:Pairing}.
%\begin{equation}
%f_{s}(l)=\frac{1}{\sqrt{6}}[\delta_{l,\vec{a}_{1}}+\delta_{l,-\vec{a}_{1}}+\delta_{l,\vec{a}_{2}}+\delta_{l,-\vec{a}_{2}}+\delta_{l,\vec{a}_{1}+\vec{a}_{2}}+\delta_{l,-\vec{a}_{1}-\vec{a}_{2}}],
%\end{equation}
%\begin{equation}
%f_{d_{xy}}(l)=\frac{1}{2\sqrt{3}}[\delta_{l,\vec{a}_{2}}+\delta_{l,-\vec{a}_{2}}-\delta_{l,\vec{a}_{1}+\vec{a}_{2}}-\delta_{l,-\vec{a}_{1}-\vec{a}_{2}}],
%\end{equation}
%\begin{equation}
%f_{d_{x^{2}-y^{2}}}(l)=\frac{1}{2}[2\delta_{l,\vec{a}_{1}}+2\delta_{l,-\vec{a}_{1}}-\delta_{l,\vec{a}_{2}}-\delta_{l,-\vec{a}_{2}}-\delta_{l,\vec{a}_{1}+\vec{a}_{2}}-\delta_{l,-\vec{a}_{1}-\vec{a}_{2}}],
%\end{equation}
%\begin{equation}
%f_{f}(l)=\frac{1}{2\sqrt{3}}[\delta_{l,\vec{a}_{1}}-\delta_{l,-\vec{a}_{1}}+\delta_{l,\vec{a}_{2}}-\delta_{l,-\vec{a}_{2}}+\delta_{l,\vec{a}_{1}+\vec{a}_{2}}-\delta_{l,-\vec{a}_{1}-\vec{a}_{2}}],
%\end{equation}
%\begin{equation}
%f_{p_{x}}(l)=\frac{1}{2}[\delta_{l,\vec{a}_{2}}-\delta_{l,-\vec{a}_{2}}+\delta_{l,\vec{a}_{1}+\vec{a}_{2}}-\delta_{l,-\vec{a}_{1}-\vec{a}_{2}}],
%\end{equation}
%\begin{equation}
%f_{p_{y}}(l)=\frac{1}{\sqrt{6}}[-2\delta_{l,\vec{a}_{1}}+2\delta_{l,-\vec{a}_{1}}+\delta_{l,\vec{a}_{2}}-\delta_{l,-\vec{a}_{2}}-\delta_{l,\vec{a}_{1}+\vec{a}_{2}}+\delta_{l,-\vec{a}_{1}-\vec{a}_{2}}].
%\end{equation}
%%%%%%%%%%%%%%%%%%%%%%%%%%%%%%%%%%%%%%%%%%%%%%%%%%%%%%%%%%%%%%%%%%%%%%%%%%%%

%%%%%%%%%%%%%%%%%%%%%%%%%%%%%%%%%%%%%%%%%%%%%%%%%%%%%%%%%%%%%%%%%%%%%%%%%%%%%
The former three are singlet pairing and the latter two are triplet case. As the triangular lattice is isotropic, the $d_{xy}$-wave and $d_{x^{2}-y^{2}}$-wave are degenerate, and the same goes for the $p_{x}$-wave and $p_{y}$-wave, here, we denote them as the $d$-wave and $p$-wave respectively\cite{Koretsune2005,Su2008}.
\begin{figure}[ptb]
\includegraphics[scale=0.45]{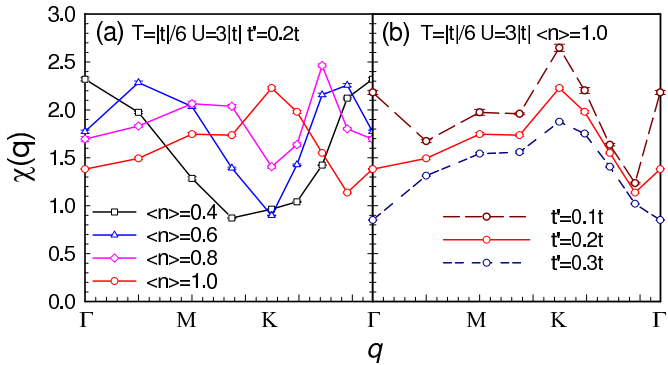}
\caption{(Color online) Spin susceptibility $\chi(q)$ versus the momentum $q$ (a) at various electron filling for $U=3|t|$, $t^{\prime}=0.2t$, $T=|t|/6$ and (b) at half filling for different $t'$.
%, where  $t'=0.1t$ ( dash dark-red line with circle),  $t'=0.2t$ (solid red line with circle), and  $t'=0.2t$ (short-dash dark blue line with circle).
Data are shown along the path $\Gamma\rightarrow$ M$\rightarrow$ K$\rightarrow\Gamma$ in the hexagonal BZ.}
\label{Fig:Xri}
\end{figure}

%%%%%%%%%%%%%%%%%%%%%%%%%%%%%%%%%%%%%%%%%%%%%%%%%%%%%%%%%%%%%%%%%%%%%%%%%%%%

\section{Results and discussion}
%%%%%%%%%%%%%%%%%%%%%%%%%%%%%%%%%%%%%%%%%%%%%%%%%%%%%%%%%%%%%%%%%%%%%%%%%%%%

%%%%%%%%%%%%%%%%%%%%%%%%%%%%%%%%%%%%%%%%%%%%%%%%%%%%%%%%%%%%%%%%%%%%%%%%%%%%%

%%%%%%%%%%%%%%%%%%%%%%%%%%%%%%%%%%%%%%%%%%%%%%%%%%%%%%%%%%%%%%%%%%%%%%%%%%%%%
%Our numerical calculations are mainly performed on lattices
%of double-48 sites with periodic boundary conditions.
By using the determinant Quantum Monte Carlo method, one author of us and his collaborators have studied the magnetic correlation of the
bilayer triangular lattice on the basis of single-band Hubbard model, % in heavily doped region, % in which ferromagnetic fluctuations are discussed in heavily doped region\cite{Hu2009}.
in which the ferromagnetic fluctuations near the van Hove singularities
were reported\cite{Hu2009}, and the doped region is $0.60 \sim 0.85$, which
corresponding to the electron filling $0.40 \sim 0.15$ in current case.
%in doped region $0.60 \sim 0.85$ %(0.15 $\leq$ $<n>$ $\leq$ 0.40)) was reported.
 %and it should be suppressed by the increasing $t'$\cite{Hu2009}.
%, where the electron filling are at least 0.5 away from half filling, namely, in heavily doped region.
%, where the doping is larger than 0.5.%when the doping was larger than 0.5. Here we study the magnetic
%fluctuation in this system around half filling to heavily doped region.
In Fig.\ref{Fig:Xri}, we present the spin susceptibility $\chi(q)$ in the electron filling region from $<n>=0.40$ to $1.0$, especially when
the system is around the half filling with different inter layer coupling terms $t'$.
Fig. \ref{Fig:Xri} (a) shows $\chi(q)$ versus the momentum $q$ at
$<n>=1.0$ (red line with circle), $<n>=0.8$ (dark line with square), $<n>=0.6$ (blue line with triangular) and $<n>=0.4$ (pink line with diamond) for  $U=3|t|$, $t^{\prime}=0.2t$ and $T=|t|/6$. At half filling, the peak of spin susceptibility is located at $K$ point. When the system is doped away from  the half filling, the peak of $\chi(q)$ moves to the $\Gamma$ point. Here, $\chi(K)$ measures the antiferromagnetic correlation and $\chi(\Gamma)$ measures the ferromagnetic fluctuations. Hence, the antiferromagnetic correlations dominate around the half filling region and ferromagnetic fluctuation dominates in the low electron fillings.  Fig. \ref{Fig:Xri} (b) shows the $\chi(q)$ with different inter layer coupling terms $t'$ at the half filling for $U=3.0|t|$ and $T=|t|/6$.
One can see that the peak of the spin susceptibility $\chi(q)$ is located at $K$ point, while $\chi(q)$ is suppressed as $t'$ increases.

\begin{figure}[tbp]
\includegraphics[scale=0.42]{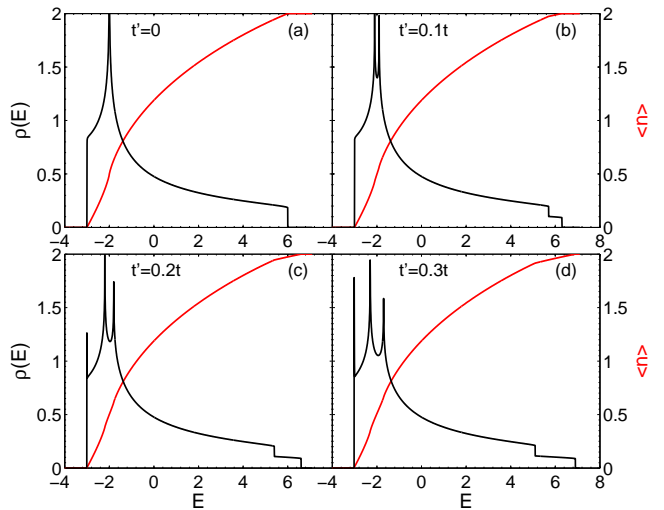}
\caption{(Color online) Dos and band fillings are functions of energy with  (a) $t'=0$, (b) $t'=0.1t$, (c) $t'=0.2t$ and (d) $t'=0.3t$, where the red lines represent
 fillings $<n>$ and the black lines represent the DOS.} \label{Fig:DOS}
\end{figure}
Such suppression, as well as the competition between ferromagnetic and antiferromagnetic correlation, could be understood from the property of the DOS in bilayer triangular lattice. %{\color{blue}
The DOS and band fillings with different $t'$ have been shown in Fig. \ref{Fig:DOS} as function of energy.%}.
One can see that, for the single-layer triangular lattice ( $t'=0$ in current case), its DOS in the non-interacting case has one van Hove singularity as the system is 0.5 doped away from the half filling.
As the inter layer coupling term $t'$ is introduced, the Van Hove singularity in the DOS tends to move further away from the half filling.
According to the itinerant electron ferromagnetic theory, the ferromagnetic fluctuations tend to the higher DOS on the Fermi surface, so ferromagnetic correlation dominates in low electron filling region.
As a result, the spin correlation at $\Gamma$ point is suppressed as the inter layer coupling term increases at half filling.

\begin{figure}[ptb]
\includegraphics[scale=0.425]{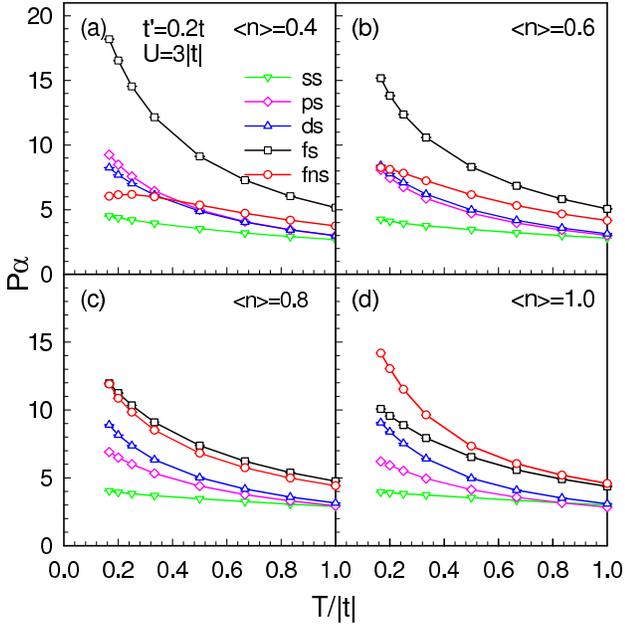}
\caption{(Color online) Pairing susceptibility for different pairing symmetries (ss: $s$ wave, ps: $p$ wave, ds: $d$ wave, fs: $f$ wave, fns: $fn$ wave) versus the temperature $T$ at $U=3|t|$ and $t^{\prime}=0.2t$. The sub-figures represent the situations of $<n>=0.4$ (a), $0.6$ (b), $0.8$ (c), and $1.0$(d) respectively.}
\label{Fig:Pairingn}
\end{figure}
%%%%%%%%%%%%%%%%%%%%%%%%%%%%%%%%%%%%%%%%%%%%%%%%%%%%%%%%%%%%%%%%%%%%%%%%%%%%%
%For the single layer triangular lattice, there is a Van Hove singularity in the density of states as the electron filling is 0.5 away from half filling.
%For a bilayer triangular lattice, the inter layer hoping term $t'$ should move such Van Hove singularity farther slightly away from the half filling. From the viewpoint of
%stoner theory, it is suggested the
%competition between ferromagnetic and antiferromagnetic correlation is governed by the presence of such Van Hove singularity, where magnetic instability might occurs\cite{Hu2009,Ma2010}.
Regarding the ferromagnetic correlation %reported in Ref.\cite{Hu2009}
and the antiferromagnetic fluctuation shown in Fig.\ref{Fig:Xri} at different electron fillings,
the competition between them indicates that paring properties in such system may also be dependent on the electron fillings.
Fig. \ref{Fig:Pairingn} presents the temperature dependence of pairing susceptibility with different symmetries at
(a) $<n>=0.4$, (b) $<n>=0.6$, (c) $<n>=0.8$ and (d) $<n>=1.0$ for $U=3|t|$ and $t^{\prime}=0.2t$.
Basically, the behaviors of paring susceptibility with all kinds of symmetry do not
change qualitatively in the major part of the temperature region we
studied. In the low electron filling region, as that shown in Fig. \ref{Fig:Pairingn} (a) and (b),
it is clear to see that the spin triplet $f$-wave, $p$-wave and the spin singlet $d$-wave pairing susceptibilities
keep growing; especially, the $f$ wave grows fastest.
The $fn$-wave and $s$-wave paring susceptibilities tend to saturate at $<n>=0.4$.
At $<n>=0.8$, Fig.\ref{Fig:Pairingn} (c) shows the $fn$ and $f$-wave paring susceptibility contest the ``race'' closely in the whole temperature region. % is complex.
As the electron filling increases up to the half filling, as that shown in Fig. \ref{Fig:Pairingn} (d), the $fn$-wave paring susceptibility tends to increase fastest, and the $d$-wave paring susceptibility also has a potential to increase faster than the $f$-wave paring susceptibility.
%{\color{blue}
However, due to the limitation of the numerical tool used here, we can not achieve arbitrarily low temperatures within the determinant Quantum Monte Carlo method, which experience the infamous fermion sign problem, and cases exponential growth in the variance of the computed results and hence an exponential growth in computer time as the lattice size is increased and the temperature is lowered.
Basically, our numerical technology works well if the electron filling is not too close to the Van Hove singularity, and in the range of $U/T\leq 36$, the error bar could be controlled within one percent for a $2 \times 48$ lattice.%}
%%%%%%%%%%%%%%%%%%%%%%%%%%%%%%%%%%%%%%%%%%%%%%%%%%%%%%%%%%%%%%%%%%%%%%%%%%%%%
\begin{figure}[ptb]
\includegraphics[scale=0.425]{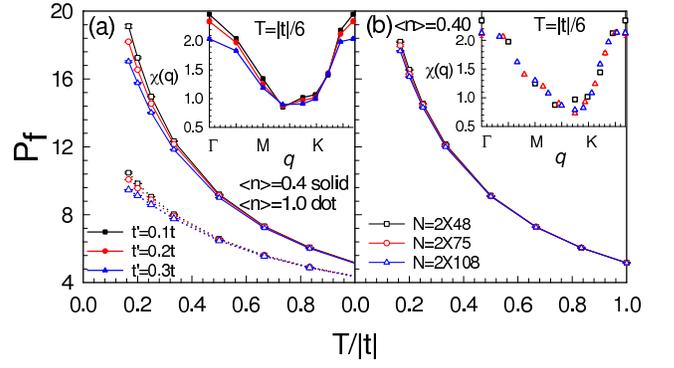}
\caption{(Color online) (a) $f$-wave pairing susceptibility at various inter layer coupling term $t^{\prime}$=0.1$t$ (dark line with square), 0.2$t$ (red line with circle) and 0.3$t$ (blue line with triangular) versus the temperature $T$ at $<n>=0.40$ (solid lines) and $<n>=1.0$ (dot lines) for $U=3|t|$. Inset: The spin susceptibility at different $t^{\prime}$ for a $2\times 48$ lattice at $T=|t|/6$, $U=3|t|$ and $<n>=0.40$.
(b) $f$-wave pairing susceptibility for a $2\times 48$ lattice, a $2\times 75$ lattice and a $2\times 108$ lattice with $t^{\prime}$=0.2$t$, $U=3|t|$ and $<n>=0.4$. Inset: The spin susceptibility for various lattices at $T=|t|/6$.}
\label{Fig:Ftp}
\end{figure}
%%%%%%%%%%%%%%%%%%%%%%%%%%%%%%%%%%%%%%%%%%%%%%%%%%%%%%%%%%%%%%%%%%%%%%%%%%%%%

Figs. \ref{Fig:Xri} and \ref{Fig:Pairingn} indicate that the competition of ferromagnetic and antiferromagnetic fluctuations in different filling region is crucial on the pairing behavior. Around half filling, the antiferromagnetic correlation
dominates in the behavior of spin correlation, and the pairing susceptibility with $fn$-wave paring symmetry is the most favorable. As the system is doped away from half filling,
the ferromagnetic correlation tends to dominate over the antiferromagnetic correlation,
and it is interesting to see that the $f$-wave pairing is the most favorable
at the ferromagnetic fluctuation dominating region,
%. Thus, the results present here indicates that the
 %ferromagnetic correlation favor the $f$-wave pairing symmetry,
 which is consistent with  previous work done by Kumar and Shastry\cite{Kumar2003}.

\begin{figure}[h]
\begin{center}
\includegraphics[scale=0.425]{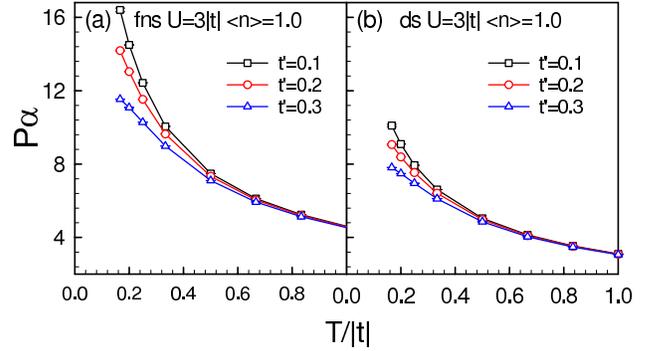}
\end{center}
\caption{(Color online) Pairing susceptibility at various inter layer coupling term $t^{\prime}$=0.1t (dark line with square), 0.2t (red line with circle) and 0.3t (blue line with triangular) versus the temperature $T$ for $U=3|t|$ and $n=1.0$. The sub-figures represent the situations of $fn$-wave (a) and $d$-wave respectively.}
\label{Fig:Dtp}
\end{figure}

%{\color{blue}
The temperature dependence of $f$-wave pairing susceptibility with different $t'$ is presented in Fig. \ref{Fig:Ftp} (a) at $<n>=0.40$ (solid lines) and  $<n>=1.0$ (dot lines).
One can see that, the inter layer hopping has little influence on the $f$-wave pairing susceptibility, whatever for $<n>=0.40$ or $<n>=1.0$.
The $t'$-dependence of $\chi(q)$ for $<n>=0.4$ are also shown in the inset of Fig. \ref{Fig:Ftp} (a). The $\chi(\Gamma)$ is suppressed very slightly as the $t'$ increases, which is consistent with
the behavior of the pairing correlation. %} {\color{blue}
In Fig.\ref{Fig:Ftp} (b), the pairing susceptibility and spin susceptibility are shown on a $2\times 48$ lattice, a $2\times 75$ lattice and a $2\times 108$ lattice for $t^{\prime}$=0.2$t$, $U=3|t|$ and $<n>=0.4$. Both the pairing susceptibility and spin susceptibility decrease slightly as the lattice size increases from $2\times 48$ to $2\times 75$, and results for $2\times 75$ and $2\times 108$ are almost the same within the error bar. Hence we may argue here that the pairing and spin susceptibility is almost independent of the lattice size. %}

%Since the fn- and f-wave pairings dominate for n>0.8 Similar to $f$-wave pairing susceptibility,
%{\color{blue}
Fig. \ref{Fig:Pairingn} shows that the $fn$- and $f$-wave pairings dominate for $<n>>$0.8 and $<n><$0.8, respectively.
The temperature dependence of $fn$-wave pairing susceptibility and $d$-wave pairing susceptibility with different $t'$ are shown in Fig. \ref{Fig:Dtp} (a) and (b) at $<n>=1.00$. %}
At half filling, one can see that the paring susceptibility is suppressed slightly by the increasing $t'$.
%As the electron filling decreases, the effect of $t'$ on the paring susceptibilities decreases. %However, the paring susceptibility is suppressed slightly by the increasing $t'$ .
This suppression is consistent with the behavior of spin susceptibility shown in Fig.\ref{Fig:Xri} in which $\chi(q)$ is also suppressed as the inter layer hoping term increases.
%As we have discussed,

%In this section we discuss the pairing correlation functions.
%Figure 8 shows the pairing correlatione of various waves versus the temperature T for $U=3$  and $t'=0.1$ . From either subfigure we note
%the following characteristics: (1) The pairing correlations of s-wave¡¢p-wave and d-wave even their temperature-variation tendency are
%similar, the pairing correlation of f-wave looks the strongest and fn-wave the weakest; (2) The pairing correlations of all waves in the
%figures decrease with temperature except for the fn-wave which is just on the opposite; (3) It is not clear whether the correlation keep
%growing or keep decreasing at lower temperatures since the sign problem prevents simulations in the low-temperature region. Comparing
%the subfigure (a) with (b) we find that all correlations at the electron filling $<n>=1.0$  are stronger than that at $<n>=0.8$  .
%%%%%%%%%%%%%%%%%%%%%%%%%%%%%%%%%%%%%%%%%%%%%%%%%%%%%%%%%%%%%%%%%%%%%%%%%%%%%

%\begin{figure}[ptb]
%\includegraphics[bb=24.5 189 536 680, width=7 cm, clip]{fig8.eps}
%\caption{(Color online). Pairing susceptibility at different electron fillings and Coulomb interactions versus the temperature $T$ for $t^{\prime}=t/10$. The sub-figures represent the situations of $d$ wave and $f$ wave. respectively.}
%\label{fig:fig8}
%\end{figure}

%%%%%%%%%%%%%%%%%%%%%%%%%%%%%%%%%%%%%%%%%%%%%%%%%%%%%%%%%%%%%%%%%%%%%%%%%%%%%

\section{Conclusion}
%%%%%%%%%%%%%%%%%%%%%%%%%%%%%%%%%%%%%%%%%%%%%%%%%%%%%%%%%%%%%%%%%%%%%%%%%%%%%
To make a summary, we have studied the magnetic and paring correlation of the
single-band Hubbard model on a bilayer triangular lattice.
We performed the determinant Quantum Monte Carlo simulations on the magnetic correlation and paring susceptibility for a variety of electron fillings, temperatures and pairing symmetries.  Around half filling, where the peak of the spin structure factor is located at $K$ point, the $fn$-wave pairing susceptibility dominates. As the electron filling decreases, the peak of spin correlation moves away from $K$, and finally locates at the $\Gamma$ point\cite{Hu2009}, which indicates the ferromagnetic fluctuation is stronger than the antiferromagnetic type  when the electron filling is low enough. And the corresponding, the $f$-wave pairing susceptibility is enhanced and the $fn$-wave paring susceptibility is suppressed as the electron filling decreases, especially at low temperature. As a result, the $f$-wave pairing susceptibility dominates as the electron filling is lower than 0.8. Moreover, both the spin correlation and paring susceptibility are suppressed by the increasing inter layer coupling $t'$. Note that our calculations only give reliable results at $T>t/6$. Therefore it is not conclusive wether the triplet $f$-wave really diverges or not as $T \rightarrow0.$ However, it would be important that there is a possibility of triplet superconductivity in the Hubbard model on a bilayer triangular lattice.
%, especially in the low density region.

\acknowledgements

This work is supported by NSFC Grant. No. 11104014,
Research Fund for the Doctoral Program of Higher Education of China
20110003120007, SRF for ROCS (SEM), and the Fundamental Research  Funds for the Central Universities in China under 2011CBA00108.

\end{document}